\documentclass[oldversion]{aa}
\usepackage{graphicx}
\usepackage{ulem}
\usepackage{threeparttable}

      \def\new#1 {{\bf #1 }}
      \def\cut#1 {\sout{#1} }

\begin{document}
\def\ffam {\hbox{$\,.\!\!^{\prime}$}}
\def\ffas {\hbox{$\,.\!\!^{\prime\prime}$}}
\def\ffM {\hbox{$\,.\!\!^{\rm M}$}}
\def\ffm {\hbox{$\,.\!\!^{\rm m}$}}
\def\ffs {\hbox{$\,.\!\!^{\rm s}$}}

\title{The density, the cosmic microwave background, and the proton-to-electron mass ratio
       in a cloud at redshift 0.9.}


\author{C.~Henkel\inst{1}
        \and
        K.~M. Menten\inst{1}
        \and
        M.~T. Murphy\inst{3}
        \and
        N. Jethava\inst{1,2}
        \and
        V.~V. Flambaum\inst{4}
        \and
        J.~A.~Braatz\inst{5}
        \and
        S. Muller\inst{6,7}
        \and
        J. Ott\inst{5,8}\thanks{J{\"u}rgen Ott is a Jansky Fellow of the National Radio Astronomy Observatory (NRAO).}
        \and
        R.~Q.~Mao\inst{9}}

\offprints{C. Henkel, \email{chenkel@mpifr-bonn.mpg.de}}

\institute{Max-Planck-Institut f\"ur Radioastronomie, Auf dem H\"ugel 69, D-53121 Bonn, Germany
           \and
           National Institute of Standards and Technology, M.S. 817.03, 325 Broadway, 80305 Boulder, USA
           \and
           Centre for Astrophysics and Supercomputing, Swinburne University, PO Box 218, Victoria 3122, Australia
           \and
           School of Physics, University of New South Wales, Sydney, N.S.W. 2052, Australia
           \and
           National Radio Astronomy Observatory, 520 Edgemont Road, Charlottesville, VA 22903, USA
           \and
           Academia Sinica Institute of Astronomy and Astrophysics, PO Box 23--141, Taipei, 106, Taiwan
           \and
           Onsala Space Observatory, SE 439--92 Onsala, Sweden
           \and
           CalTech, 1200 E. California Blvd., Caltech Astronomy, 105--24, Pasadena, CA 91125--2400, USA
           \and
           Purple Mountain Observatory, Chinese Academy of Sciences, 210008 Nanjing, China}

\date{Received date / Accepted date}

\abstract
{Based on measurements with the Effelsberg 100-m telescope, a multi-line study of molecular species is presented
toward the gravitational lens system PKS\,1830--211, which is by far the best known target to study dense cool gas 
in absorption at intermediate redshift. Determining average radial velocities and performing Large Velocity Gradient 
radiative transfer calculations, the aims of this study are (1) to determine the density of the gas, (2) to constrain 
the temperature of the cosmic microwave background (CMB), and (3) to evaluate the proton-to-electron mass ratio at 
redshift $z$ $\sim$ 0.89. Analyzing data from six rotational HC$_3$N transitions (this includes the $J$=7$\leftarrow$6 
line, which is likely detected for the first time in the interstellar medium) we obtain $n$(H$_2$) $\sim$ 2600\,cm$^{-3}$ 
for the gas density of the south-western absorption component, assuming a background source covering factor, which 
is independent of frequency. With a possibly more realistic frequency dependence proportional to $\nu^{0.5}$ (the maximal 
exponent permitted by observational boundary conditions), $n$(H$_2$) $\sim$ 1700\,cm$^{-3}$. Again toward the south-western 
source, excitation temperatures of molecular species with optically thin lines and higher rotational constants are, 
on average, consistent with the expected temperature of the cosmic microwave background, $T_{\rm CMB}$ = 5.14\,K. 
However, individually, there is a surprisingly large scatter which far surpasses expected uncertainties. A comparison 
of CS $J$=1$\leftarrow$0 and 4$\leftarrow$3 optical depths toward the weaker north-western absorption component 
results in $T_{\rm ex}$ = 11\,K and a 1-$\sigma$ error of 3\,K. For the main component, a comparison of velocities 
determined from ten optically thin NH$_3$ inversion lines with those from five optically thin rotational transitions 
of HC$_3$N, observed at similar frequencies, constrains potential variations of the proton-to-electron mass ratio $\mu$ 
to $\Delta\mu$/$\mu$ $<$1.4$\times$10$^{-6}$ with 3-$\sigma$ confidence. Also including optically thin rotational lines 
from other molecular species, it is emphasized that systematic errors are $\Delta V$ $<$ 1\,km\,s$^{-1}$, corresponding 
to $\Delta\mu$/$\mu$ $<$ 1.0$\times$10$^{-6}$.}
{}

\titlerunning{Molecular absorption lines in PKS\,1830--211}

\authorrunning{Henkel et al.}

\maketitle


\section{Introduction}

In the Galaxy, detailed studies of molecular clouds and complexes can be performed at subparsec resolution 
with a large number of spectroscopic tools in a large variety of regions. However, even interferometric data 
from nearby galaxies outside the Local Group only provide typical angular resolutions of a few 10\,pc for 
quasi-thermal line emission, matching the size of entire giant molecular clouds. Studies of quasi-thermal line 
emission from even more distant targets are hampered by small beam filling factors and low flux densities. 
To circumvent these problems, one can search for masers which may be strong enough to be seen out to cosmological 
distances (Impellizzeri et al. 2008). Similarly promising is the study of absorption lines. In this case the 
``effective beamwidth'' is determined by the size of the background continuum source, which may be much more 
compact than the main beam of a single-dish telescope or the synthesized beam of an interferometer system. 
It is then the strength of the background continuum, which determines the sensitivity of the observations. 

In the following we will discuss spectroscopic data obtained toward a particularly prominent radio source, 
which is observed at an intermediate redshift ($z$$\sim$0.89), {\it (1) to explore the physical properties of a 
distant molecular cloud, (2) to use these properties for a test of the correlation between $T_{\rm CMB}$ and 
redshift}, and (3) to {\it search for variations of the proton-to-electron mass ratio as a function of time}. 

So far, only a few studies exist of individual giant molecular clouds at intermediate redshift (see, Wiklind \& 
Combes 1994, 1995, 1996a, 1996b; Kanekar et al. 2005). A good knowledge of their physical and chemical conditions 
is, however, mandatory for an evaluation of systematic differences between clouds at small and large cosmological 
distances.  

According to the standard model of cosmology, the cosmic microwave background (CMB) should follow a Planck curve 
of temperature $T_{\rm CMB}$ = (2.725$\pm$0.001) $\times$ (1+$z$), with 0 $\leq$ $z$ $\leq$ 1088$^{+1}_{-2}$ 
denoting redshift (e.g., Battistelli et al. 2002; Fixsen \& Mather 2002; Spergel et al. 2003). Determining deviations 
from this relation would be a powerful tool to challenge the standard model of cosmology, highlighting possible 
mechanisms acting upon the photons of the cosmic microwave background (e.g., Lima et al. 2000). The best 
experimental method to measure $T_{\rm CMB}$(1+$z$) is a multi-spectral line analysis of ions, atoms or molecules 
that experience negligible excitation by particle collisions and by radiation from nearby sources, thus being almost 
exclusively excited by the isotropic radiation of the CMB. A comparison of line temperatures then yields excitation 
temperatures, which provide firm limits to $T_{\rm CMB}$. 

According to the standard model of particle physics, fundamental constants, at least in their low-energy limits, 
should be independent of time and location. The fundamental constants of physics and astronomy are well defined locally. 
Although there have been many detailed studies searching for spatial or temporal variations, none have been 
convincingly established (e.g., Uzan 2003; Garc\'{\i}a-Berro et al. 2007; Flambaum 2008). However, this ``constancy 
of constants'' may not necessarily hold over the largest spatial and temporal scales, which are inaccessible by geological 
or astronomical studies of nearby targets. Now, advances in observational sensitivity make it possible to measure fundamental 
constants for sources in the distant Universe, viewed at times billions of years ago and preceding the formation of the solar 
system (e.g., Bahcall et al. 1967; Varshalovich \& Levshakov 1993; Cowie \& Songaila 1995; Webb et al. 1999). 

When searching for variations of physical constants during the last $\sim$10$^{10}$\,yr, the most frequently studied parameter
is the fine-structure constant $\alpha$ (Webb et al. 1999, 2001; Murphy et al. 2003, 2008b; Chand et al.  2004; Levshakov 
et al. 2006, 2007; Rosenband et al. 2008). Composite parameters consisting of $\alpha$, the proton-to-electron mass ratio 
$\mu$, and the proton-g factor $g_{\rm p}$ have also been analyzed (Drinkwater et al. 1998; Carilli et al. 2000; Murphy et 
al. 2001; Chengalur \& Kanekar 2003; Kanekar et al. 2004, 2005; Tzanavaris et al. 2005, 2007). It may well be possible 
that the chromodynamic scale varies faster than the corresponding scale of quantum-electrodynamics and that, as a consequence, 
the proton-to-electron mass ratio $\mu$ may show larger deviations from local values than the fine-structure constant (e.g., 
Calmet \& Fritzsch 2002; Langacker et al. 2002; Flambaum et al. 2004) though this is highly model dependent (Dent 2008). 

To date, studies focusing specifically on potential variations of $\mu$ are rare, yielding 1$\sigma$ confidence levels of 
order $\Delta$$\mu$/$\mu$ $\sim$ 10$^{-5}$ (Levshakov et al. 2002; Ivanchik et al. 2005; Reinhold et al. 2006; King et al. 
2008). Different powers of $\mu$ define the scales of electronic, vibrational, and rotational intervals in molecular spectra 
so that ro-vibronic transitions of the H$_2$ molecule, observed toward distant quasars, could be used. 

An alternative and potentially even more precise approach is the analysis of the inversion line spectrum of ammonia (NH$_3$). 
Variations of $\mu$ could be identified as shifts in the relative frequencies of the inversion versus those of the rotational 
lines. After NH$_3$ had been observed for the first time at significant redshift, in B0218+357 at $z$=0.685 (Henkel et al. 2005), 
Flambaum \& Kozlov (2007) compared the frequencies of the three detected inversion lines with those of rotational lines of CO, 
HCO$^+$, and HCN (Wiklind \& Combes 1995; Combes \& Wiklind 1995). This resulted in $|\Delta$$\mu|$/$\mu$ = (0.6$\pm$1.9) $\times$ 
10$^{-6}$. Even more recently, Murphy et al. (2008a) derived from a more thorough analysis of the ammonia spectra and newly 
observed HCO$^+$ and HCN data $\Delta \mu$/$\mu$ = (0.74$\pm$0.47$_{\rm stat}$$\pm$0.76$_{\rm sys}$) $\times$ 10$^{-6}$. As 
a cautionary note it has to be emphasized, however, that the NH$_3$ inversion and HCO$^+$ and HCN rotational lines have frequencies, 
which differ by a factor of about ten, while the morphology of the background continuum is frequency dependent (e.g., Jethava 
et al. 2007). B0218+357 is located at a lookback time of 6.0\,Gyr, adopting a standard $\Lambda$-cosmology with $H_0$ = 
73\,km\,s$^{-1}$\,Mpc$^{-1}$, $\Omega_{\rm m}$ = 0.28, and $\Omega_{\Lambda}$ = 0.72 (Spergel et al. 2007). 

Among the redshifted sources showing molecular absorption spectra, the gravitational lens PKS\,1830--211 stands out and has
distinct advantages over B0218+357 with respect to line strengths and chemical complexity. With $z$ = 0.88582 for its main
absorption feature and a lookback time of 7.0\,Gyr, it is also the most distant of these sources presently known. With a 
lensed $z$=2.5 quasar providing an exceptionally bright radio background, a vast number of molecular absorption lines could 
be detected (Wiklind \& Combes 1996b, 1998; Gerin et al. 1997; Chengalur et al. 1999; Menten et al. 1999; Muller et al. 2006). A 
total of 11 NH$_3$ inversion lines has been observed toward the source (Henkel et al. 2008; J.  Ott, priv. comm.). With the 
kinetic temperature obtained from the NH$_3$ data, here we provide the complementary estimate of the density of the gas as 
well as excitation temperatures, which can be compared with the expected temperature of the CMB. To determine $\Delta\mu$/$\mu$, 
frequencies of NH$_3$ inversion lines will be compared with newly measured optically thin rotational lines also detected in 
the $\lambda$ = 0.7--2.5\,cm wavelength range.

\section{Observations}

The observations were carried out with the primary focus $\lambda$ = 1.9, 1.3, 1.0, and 0.7\,cm receivers of the 
100-m telescope at Effelsberg/Germany\footnote{The 100-m telescope at Effelsberg is operated by the Max-Planck-Institut 
f{\"u}r Radioastronomie (MPIfR) on behalf of the Max-Planck-Gesellschaft (MPG).} between August 2001 and March 2002. For 
full width to half power (FWHP) beamwidths, system temperatures, and aperture efficiencies, see Table~1. The backend 
was an 8192 channel autocorrelator, which was split into eight segments, covering selected frequency ranges within a total 
bandwidth of 500\,MHz. Each of the eight segments had a bandwidth of 40\,MHz with 512 channels or, at 0.7\,cm, 80\,MHz with 
256 channels, yielding channel spacings of 0.7--2.0\,km\,s$^{-1}$. At 1.3 and 0.7\,cm, dual channel receivers were used. 
Observing with these receivers, therefore at least two of the eight segments were centered at the same frequency to maximize 
sensitivity by observing simultaneously the two orthogonal linear polarizations. 

As already mentioned (Sect.\,1), PKS\,1830--211 is one of the strongest compact continuum sources in the radio sky. Therefore, 
pointing could be checked toward the source itself and was found to be accurate to about 5\arcsec. The continuum was also 
used to calibrate the spectral lines to obtain line-to-continuum flux density ratios, which could be determined with high
accuracy (see Sect.\,4).

\begin{table}
\label{tab1}
\begin{threeparttable}
\caption[]{Observational parameters.}
\begin{flushleft}
\begin{tabular}{ccccc}
\hline 
     $\nu$ & $\lambda$ & $\theta_{\rm b}^{\rm a)}$ & $T_{\rm sys}^{\rm b)}$ & $\eta_{\rm A}^{\rm c)}$ \\
     (GHz) &   (cm)    &           ($''$)          &     (K)                &                         \\
\hline
           &           &                           &                        &                         \\
13.5--18.7 &   1.9     &  60--45                   &      40                &   0.39--0.32            \\
18.0--26.0 &   1.3     &  45--35                   &      70                &   0.33--0.26            \\
27.0--36.7 &   1.0     &  32--23                   &      75                &      0.30               \\
41.1--49.7 &   0.7     &  23--19                   &     150                &   0.36--0.13            \\
           &           &                           &                        &                         \\
\hline
\end{tabular}
\begin{tablenotes}
\item[a)] Full Width to Half Power (FWHP) beam width in arcsec. 
\item[b)] System temperatures in units of antenna temperature ($T_{\rm A}^*$). 
\item[c)] Aperture efficiency.
\end{tablenotes}
\end{flushleft}
\end{threeparttable}
\end{table}

\section{Results}

PKS\,1830--211 is composed of a background continuum source at redshift $z$=2.507, possibly a blazar, and a lensing face-on 
spiral galaxy at $z$=0.88582 along its line-of-sight (Wiklind \& Combes 1996b; Lidman et al. 1999; Courbin et al. 2002; 
Winn et al. 2002; de Rosa et al. 2005). The lens splits the background continuum into three main components, two compact hotspots, 
a north-western and a south-eastern one separated by $\sim$1\arcsec, and an Einstein ring, which is prominent at low ($\ll$10\,GHz) 
frequencies. The bulk of the molecular absorption is seen toward the south-western hotspot (e.g., Frye et al. 1997; Swift et al. 
2001). 

\begin{figure}[t]
\vspace{-0.0cm}
\centering
\resizebox{19.0cm}{!}{\rotatebox[origin=br]{-90}{\includegraphics{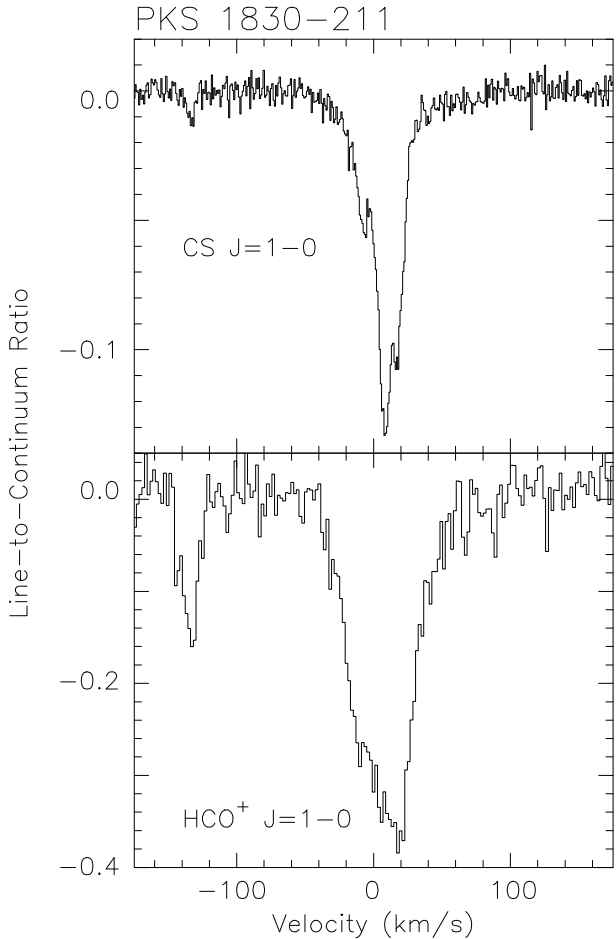}}}
\vspace{-0.7cm}
\caption{The CS and HCO$^+$ $J$=1$\leftarrow$0 absorption lines with a Local Standard of Rest (LSR) velocity scale relative to 
$z$=0.88582, observed toward PKS\,1830--211. $V_{\rm LSR}$ = $V_{\rm HEL}$ + 12.4\,km\,s$^{-1}$. Channel spacings are 0.9 and 
2.0\,km\,s$^{-1}$, respectively. The continuum level accounts for the entire source. Due to a highly frequency dependent noise 
diode signal at 47.294\,GHz, the calibration of the HCO$^+$ line is uncertain.
\label{fig1}}
\end{figure}

\begin{figure}[t]
\vspace{0.0cm}
\centering
\resizebox{19.0cm}{!}{\rotatebox[origin=br]{-90}{\includegraphics{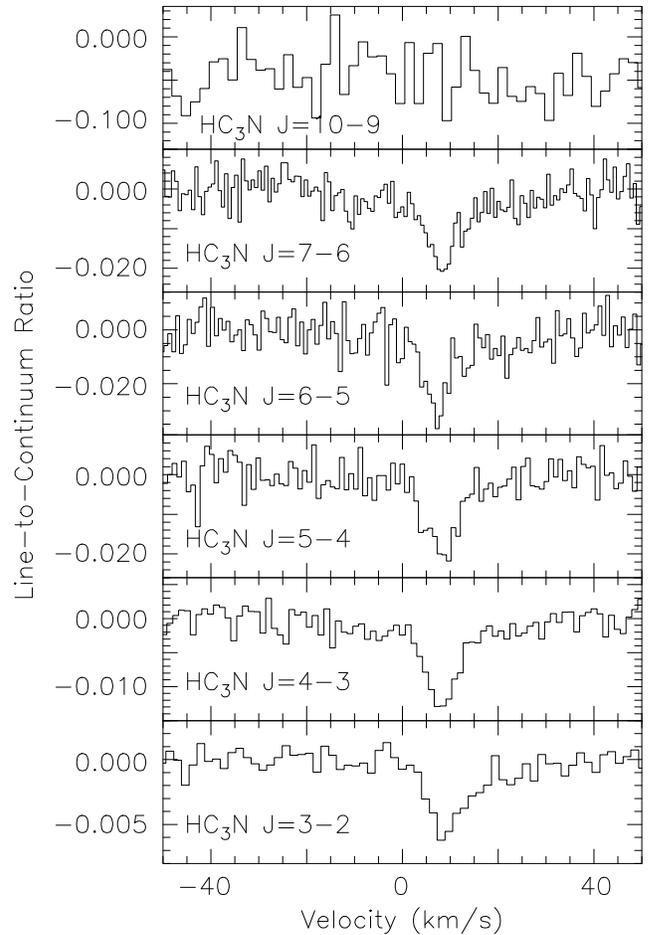}}}
\vspace{-0.7cm}
\caption{HC$_3$N absorption lines with a Local Standard of Rest (LSR) velocity scale relative to $z$=0.88582, observed toward 
PKS\,1830--211. Channel spacings are, from top to bottom, 1.9, 0.7, 0.8, 1.0, 1.2, and 1.6\,km\,s$^{-1}$. The continuum level
accounts for the entire source.
\label{fig2}}
\end{figure}

\begin{figure}[t]
\vspace{0.0cm}
\centering
\resizebox{19.0cm}{!}{\rotatebox[origin=br]{-90}{\includegraphics{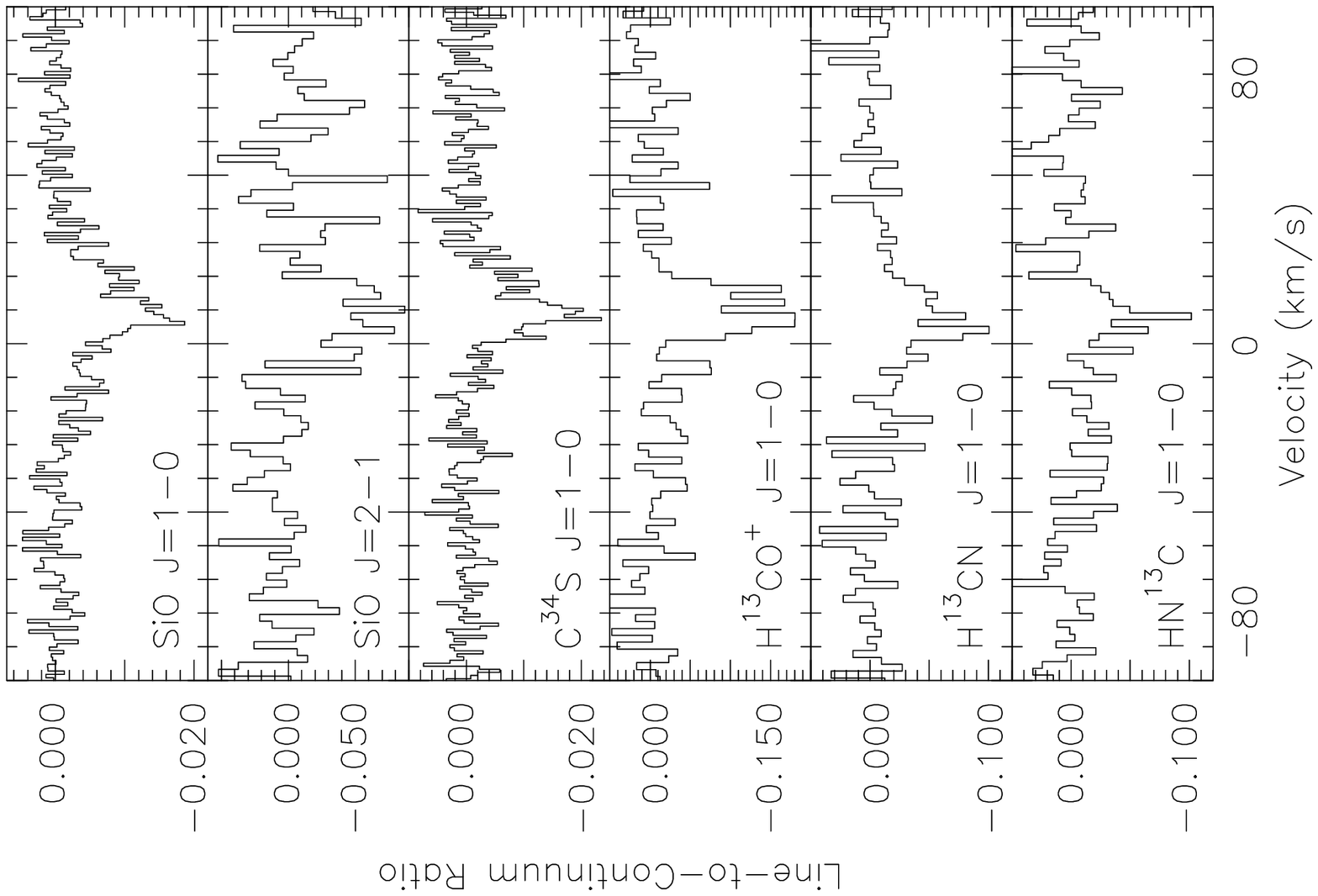}}}
\vspace{-0.7cm}
\caption{Additional optically thin absorption lines with a Local Standard of Rest (LSR) velocity scale relative to $z$=0.88582, observed 
toward PKS\,1830--211. Channel spacings are, from top to bottom, 1.0, 2.0, 0.9, 2.0, 2.0, and 2.0\,km\,s$^{-1}$. The adopted continuum 
level includes the entire source.
\label{fig3}}
\end{figure}

Figures~\ref{fig1}--\ref{fig3} show the line profiles measured at Effelsberg with the ordinate displaying absorption in units 
of the observed continuum flux density. No significant indications for variable lineshapes were found during the time interval 
of the observations (Sect.\,2), which is consistent with the optical depth monitoring of Muller \& Gu{\'e}lin (2008; their 
Fig.\,5). Since about one third of the total continuum emission is absorbed by the foreground molecular cloud at 3\,mm 
wavelength (e.g., Frye et al. 1997; Wiklind \& Combes 1998; Swift et al. 2001), the upper panel of Fig.\,\ref{fig1} may 
show a CS profile with intermediate ($\tau$ $\sim$ 1) peak optical depth. 

Ignoring the blueshifted feature at $V_{\rm LSR}$$\sim$--135\,km\,s$^{-1}$, the only spectral component known to absorb
the north-eastern continuum hotspot (see, e.g., Sect. 4.1.4 and Wiklind \& Combes 1998; Muller et al. 2006), this CS 
$J$=1$\leftarrow$0 profile may be decomposed into four velocity components. Gaussian fitting reveals a main feature at $V_{\rm LSR}$ 
= (8$\pm$1)\,km\,s$^{-1}$, a slightly weaker redshifted one at (18$\pm$1)\,km\,s$^{-1}$, a weak narrow blueshifted feature at 
\hbox{(--8$\pm$1)\,km\,s$^{-1}$}, and a similarly weak but broad line wing encompassing several tens of km\,s$^{-1}$. The lineshape 
roughly matches that of the opacity profiles presented by Muller et al. (2006; their Fig.\,5), also indicating (as the line-to-continuum 
flux density ratio) that the line is not fully saturated. We note, however, that the CS intensity ratio between the main  
($\sim$8\,km\,s$^{-1}$) and the redshifted component ($\sim$18\,km\,s$^{-1}$) is slightly smaller than for the HNC and 
H$^{13}$CO$^+$ $J$=2$\leftarrow$1 transitions; it is much smaller than for the HC$^{18}$O$^+$ $J$=2$\leftarrow$1 line, which must be 
optically thin. Muller et al. (2006) give peak optical depths of $\tau$ = 1.65, 0.27, and 0.15 for these three lines, respectively 
(their Table~4). The possibility that CS $J$=1$\leftarrow$0 is moderately optically thick ($\tau$$\sim$1) highlights the possibility 
that the continuum source covering factor may be smaller at 26\,GHz than at 3\,mm (see Sect.\,4.1). Interestingly, the $J$=1$\leftarrow$0 
peak line-to-continuum ratios of CS (Fig.\,\ref{fig1}) and C$^{34}$S (Fig.\,\ref{fig3}) differ by a factor of $\sim$10, consistent 
with the corresponding ratio derived by Muller et al. (2006; their Table~7) for the $J$=4$\leftarrow$3 transition. 

The HCO$^+$ $J$=1$\leftarrow$0 profile (Fig.\,\ref{fig1}, lower panel), taken at a time when the $J$=2$\leftarrow$1 opacity was 
particularly high (Muller \& Gu{\'e}lin 2008), looks quite different. Here even the line wings at velocities only showing weak 
CS absorption are quite pronounced and the central three velocity components seen in CS $J$=1$\leftarrow$0 appear to have merged. 
Remarkably, the velocity of the HCO$^+$ absorption peak lies near +20\,km\,s$^{-1}$, well offset from the peak of the CS 
$J$=1$\leftarrow$0 absorption. The properties of the HCO$^+$ line may either be caused by a very large optical depth or by
tracing a different volume of gas. Since the $J$=2$\leftarrow$1 lines of the rare HCO$^+$ isotopologues show a ``normal'' optical 
depth profile (Muller et al. 2006), a high optical depth is the preferred explanation.

Figure~\ref{fig2} shows five detected HC$_3$N profiles. The lines do not absorb more than a few percent of the total continuum and all 
peak near 8\,km\,s$^{-1}$. From the $J$=3$\leftarrow$2 to the 6$\leftarrow$5 transition, absolute peak line intensities rise from about 
0.5\% to 3.5\% of the background continuum level, only to decrease further up the $J$--ladder. To our knowledge, it is the first time 
the HC$_3$N $J$=7$\leftarrow$6 line has been detected in the interstellar medium. With its rest frequency of 63.686\,GHz, it is absorbed 
by atmospheric O$_2$ in more nearby sources. 

Figure~\ref{fig3} shows additional spectral features which also absorb only a tiny fraction of the continuum background. These include 
transitions of SiO and rare isotopic species of CS, HCN, HCO$^+$, and HNC.

\section{Discussion}

\subsection{Excitation analysis}

Before studying molecular excitation in detail, it is important to obtain an estimate of the interplay between the radio continuum
background and the intervening molecular cloud. PKS\,1830--211 is a variable source (e.g., Jin et al. 2003). However, for the 
limited time interval of our observations we find the depths of the absorption lines and the continuum flux densities to be correlated. 
This leaves the essential parameter, the line-to-continuum ratio, sufficiently unaffected to combine spectra observed at slighty different 
times (see also Wiklind \& Combes 1998; Muller et al. 2006; Henkel et al. 2008; Muller \& Gu{\'e}lin 2008). A more intricate problem is 
the background source coverage factor, which may vary as a function of frequency. Often, the size of the continuum background decreases 
with frequency (e.g., Lobanov 1998), while the size of the molecular cloud does not vary. If the cloud is located along the line-of-sight 
to the center of the continuum source, higher frequencies may yield higher source covering factors, while the opposite may hold when the 
cloud is only absorbing the periphery of the radio background.

At $\nu$ $\ga$ 8\,GHz, the absorbed south-western continuum source of PKS\,1830--211 (see Sect.\,1) contributes (38$\pm$2)\% to the total 
radio flux density (e.g., Nair et al. 1993; van Ommen et al. 1995; Carilli et al. 1998; Muller et al. 2006; Menten et al. 2008). According 
to Frye et al.  (1997), the molecular cloud coverage of this continuum component is $\sim$70\%. However, Wiklind \& Combes (1998) and Carilli 
et al. (1998) find values consistent with 100\%, which is corroborated by the more recent monitoring results of Muller \& Gu{\'e}lin (2008). 
This indicates that at a frequency of 94\,GHz, the covering factor of the total continuum is indeed $f_{\rm c}$ = 0.38$\pm$0.02. With the 
CS $J$=1$\leftarrow$0 line, redshifted to 26\,GHz and absorbing the continuum at a (13$\pm$1)\% level (Sect.\,3 and Fig.\,\ref{fig1}), the 
variation of $f_{\rm c}$ between 26 and 94\,GHz is then limited to $f_{\rm c}$ $\propto$ $\nu^{0.0...0.8}$. The upper limit to the exponent 
would only hold, if the CS $J$=1$\leftarrow$0 line were fully saturated. Because this is certainly not the case (see Sect.\,3), the actual 
maximal exponent must be smaller. For $\tau$(CS $J$=1$\leftarrow$0) = 1, a reasonable guess implying $f_{\rm c}$ $\sim$ 0.2 at 26\,GHz, the 
maximal exponent becomes 0.5. A negative exponent for $f_{\rm c}$ as a function of frequency can be excluded because there is no observational 
evidence for larger source covering factors at lower frequencies (for OH and H{\sc i}, see Chengalur et al.  1999; Koopmans \& de Bruyn 2005) 
and because then the extended, arcsecond sized Einstein ring becomes prominent (e.g., Jauncey et al. 1991). 

With five detected and one undetected rotational transition, we can carry out the first multilevel HC$_3$N study of a significantly 
redshifted source (see Mauersberger et al. 1990 for an analogous study of a nearby starburst galaxy). With its small rotational 
constant, HC$_3$N is particularly useful to constrain molecular excitation along the line-of-sight to PKS\,1830--211. Two SiO transitions 
and comparisons of our data with spectra observed at higher frequencies provide additional constraints. The analysis is based on measured 
``apparent'' optical depths,
$$
\tau_{\rm app} = -{\rm ln}\left(1 - \frac{|T_{\rm L}|}{T_{\rm C}} \right),
$$
with $T_{\rm L}$ and $T_{\rm C}$ denoting line and continuum temperature and with $\tau_{\rm app}$ $\sim$ $|T_{\rm L}|$/$T_{\rm C}$
being a good approximation in the case of optically thin lines (beware of the typo in the corresponding equation of Henkel et al. 2008).

\begin{table}
\label{tab2}
\begin{threeparttable}
\caption[]{Line widths and optical depths of HC$_3$N transitions (see Sect.\,4.1.1)$^{\rm a)}$.}
\begin{flushleft}
\begin{tabular}{cccc}
\hline 
  Line                         &      FWHP      & $\tau_{\rm peak}$ & $\int{\tau {\rm d}V}$ \\
\hline
                               &                &                   &                       \\
HC$_3$N $J$=3$\leftarrow$2     & 10.6$\pm$1.2   & 0.0057$\pm$0.0008 &    0.060$\pm$0.005    \\
HC$_3$N $J$=4$\leftarrow$3     &  9.3$\pm$1.0   & 0.0131$\pm$0.0018 &    0.121$\pm$0.010    \\
HC$_3$N $J$=5$\leftarrow$4     &  8.6$\pm$1.0   & 0.0211$\pm$0.0031 &    0.181$\pm$0.017    \\
HC$_3$N $J$=6$\leftarrow$5     &  9.0$\pm$1.9   & 0.0280$\pm$0.0069 &    0.251$\pm$0.034    \\
HC$_3$N $J$=7$\leftarrow$6     & 12.2$\pm$2.1   & 0.0173$\pm$0.0035 &    0.210$\pm$0.024    \\
HC$_3$N $J$=10$\leftarrow$9    &     --         &       --          &        $<$0.235       \\
                               &                &                   &                       \\
\hline
\end{tabular}
\begin{tablenotes}
\item[a)] Full Width to Half Power (FWHP) line widths, peak apparent optical depths, and integrated 
optical depths obtained from one component Gaussian fits. The upper limit to the HC$_3$N 
$J$=10$\leftarrow$9 integrated opacity is a 3$\sigma$ value. 
\end{tablenotes}
\end{flushleft}
\end{threeparttable}
\end{table}

\subsubsection{HC$_3$N}

To reproduce the HC$_3$N data (Fig.\,\ref{fig2}), we use a Large Velocity Gradient (LVG) model including 22 levels up to 92\,K above 
the ground state. Collision rates were taken from Green \& Chapman (1978). The bulk of the ammonia (NH$_3$) column density arises
from gas at $T_{\rm kin}$ $\sim$ 80\,K (Henkel et al. 2008). This value was adopted. As already mentioned (Sect.\,3), the HC$_3$N lines 
only absorb a few percent of the total continuum background (Fig.\,\ref{fig2}; Table~2). This is similar to the NH$_3$ lines also seen 
at cm-wavelengths. Since the NH$_3$ lines are definitely optically thin (the hyperfine satellites are at best barely visible), since 
HC$_3$N and NH$_3$ frequencies are similar, and since the continuum source covering factor is of order $f_{\rm c}$ = 0.10--0.38 
(Fig.\,\ref{fig1} and Sect.\,4.1), we conclude that all detected HC$_3$N lines must also be optically thin. 

For the LVG code, a spherical cloud geometry was assumed, but with the lines being optically thin, results do not depend on cloud 
morphology. We consider the integrated optical depths to be more reliable than the peak optical depths. For the input parameters we 
therefore only adopted the peak optical depth of the very well fitted $J$=5$\leftarrow$4 line, accounting for a continuum source covering 
factor $f_{\rm c}$ = 0.3 and using the ratios of the integrated optical depths to obtain the peak optical depths of the other lines. This 
results, ignoring potential uncertainties in the continuum coverage, which are difficult to quantify, in the peak optical depths
displayed in the central column of Table~3.

\begin{table}
\label{tab3}
\caption[]{HC$_3$N peak optical depths corrected for incomplete continuum source coverage (see Sect.\,4.1.1). $f_{\rm c}$ is the source
covering factor, $\nu$ denotes the redshifted frequency.}
\begin{flushleft}
\begin{tabular}{ccc}
\hline 
  Line                         & $f_{\rm c}$ = 0.3  & $f_{\rm c}$= 0.2$\times$($\nu$/26\,GHz)$^{0.5}$ \\
\hline
                               &                    &                    \\
HC$_3$N $J$=3$\leftarrow$2     & 0.0239$\pm$0.0046  &  0.0462$\pm$0.0089 \\
HC$_3$N $J$=4$\leftarrow$3     & 0.0482$\pm$0.0093  &  0.0807$\pm$0.0156 \\
HC$_3$N $J$=5$\leftarrow$4     & 0.0721$\pm$0.0106  &  0.1084$\pm$0.0158 \\
HC$_3$N $J$=6$\leftarrow$5     & 0.1000$\pm$0.0221  &  0.1367$\pm$0.0302 \\
HC$_3$N $J$=7$\leftarrow$6     & 0.0837$\pm$0.0174  &  0.1059$\pm$0.0219 \\
                               &                    &                    \\
\hline
\end{tabular}
\end{flushleft}
\end{table}

As a first approach we neglect collisional excitation and only vary the temperature of the cosmic microwave background to simulate 
the observational data with minimized reduced $\chi^2$ values, $\chi^2_{\rm red}$ = $\chi^2$/($N$--$P$) ($N$=5: number of lines, 
$P$=2: number of free parameters; see Jethava et al. 2007). At a redshift of $z$=0.88582, we expect $T_{\rm CMB}$ = (2.725$\pm$0.001) 
$\times$ (1+$z$) = (5.139$\pm$0.002)\,K. With this value, the minimal $\chi^2_{\rm red}$ is an unacceptable 8.1 (Table~4). 
Modifying $T_{\rm CMB}$, but still only permitting excitation by the cosmic microwave background, the optimal value is obtained with 
$\chi^2_{\rm red}$ = 0.8 at $T_{\rm CMB}$ $\sim$ 20\,K (model A7). Such a high $T_{\rm CMB}$ value is also unacceptable. We therefore 
have to conclude that our model assumptions are too simple. 

As the next step to higher complexity, we therefore also include collisions with H$_2$. Keeping $T_{\rm CMB}$ fixed at 5.14, we 
calculate the radiative transfer for 500\,cm$^{-3}$ $<$ $n$(H$_2$) $<$ 5000\,cm$^{-3}$. Already at $n$(H$_2$) = 500\,cm$^{-3}$, 
the quality of the fit is clearly improved (model B2 of Table~4). The optimum is reached at $n$(H$_2$) $\sim$ 2600\,cm$^{-3}$
(models B7/B8) with $\chi^2_{\rm red}$ = 0.4. In this case, calculated and measured optical depths differ by about 15\%. 
Since the opacity ratios do not vary with column density as long as the lines are optically thin, this result is not sensitive 
to the specific choice of $f_c$ ($f_{\rm c}$ $\propto$ $\tau^{-1}$ $\propto$ $N_{\rm HC_3N}^{-1}$). Already at $n$(H$_2$) = 
500\,cm$^{-3}$, the HC$_3$N $J$=1$\leftarrow$0 line is inverted. At $n$(H$_2$) = 2000\,cm$^{-3}$, the 2$\leftarrow$1 line is also 
inverted and at $n$(H$_2$) = 7000\,cm$^{-3}$, the three ground rotational lines of HC$_3$N are all inverted. Thus, at $n$(H$_2$) 
$\ga$ 7000\,cm$^{-3}$, we do not expect to obtain a detectable absorption signal in the $J$=3$\leftarrow$2 transition.

\begin{table}
\label{tab4}
\begin{threeparttable}
\caption[]{HC$_3$N LVG models and resulting reduced $\chi^2$ values (see Sect.\,4.1.1)$^{a}$.}
\begin{flushleft}
\begin{tabular}{lrrrr}
\hline 
 Model   & $T_{\rm CMB}$   & $n$(H$_2$)   &            \multicolumn{2}{c}{$\chi^2_{\rm red}$}                 \\
         &                 &              &                      \\   
         &                 &              & $f_{\rm c}$$\propto$$\nu^{0.0}$ & $f_{\rm c}$$\propto$$\nu^{0.5}$ \\   
         &       (K)       & (cm$^{-3}$)  &                                 &                                 \\   
\hline
         &                 &              &                                 &                                  \\
 A1      &     5.14        &      0       &        8.1                      &       4.1                        \\
 A2      &     6.00        &      0       &        6.0                      &       2.4                        \\
 A3      &     8.00        &      0       &        3.3                      &       0.8                        \\
 A4      &    10.00        &      0       &        2.0                      &       0.5                        \\
 A5      &    12.00        &      0       &        1.4                      &       0.6                        \\
 A6      &    15.00        &      0       &        1.0                      &       1.0                        \\
 A7      &    20.00        &      0       &        0.8                      &       1.7                        \\
 A8      &    25.00        &      0       &        0.9                      &       2.2                        \\
 A9      &    30.00        &      0       &        1.0                      &       2.6                        \\
 A10     &    50.00        &      0       &        1.4                      &       3.6                        \\
         &                 &              &                                 &                                  \\
 B1      &     5.14        &      0       &        8.1                      &       4.1                        \\
 B2      &     5.14        &    500       &        5.4                      &       2.1                        \\
 B3      &     5.14        &   1000       &        3.3                      &       0.8                        \\
 B4      &     5.14        &   1500       &        1.7                      &       0.3                        \\
 B5      &     5.14        &   1700       &        1.0                      &       0.2                        \\  
 B6      &     5.14        &   2000       &        0.8                      &       0.3                        \\  
 B7      &     5.14        &   2500       &        0.4                      &       0.7                        \\  
 B8      &     5.14        &   2700       &        0.4                      &       1.0                        \\
 B9      &     5.14        &   3000       &        0.5                      &       1.6                        \\  
 B10     &     5.14        &   4000       &        2.1                      &       4.0                        \\  
 B11     &     5.14        &   5000       &        4.9                      &       7.0                        \\  
\hline
\end{tabular}
\begin{tablenotes}
\item[a)] Assumed kinetic temperature: $T_{\rm kin}$ = 80\,K. For the reduced $\chi^2$ values with the frequency independent 
continuum source coverage factor $f_{\rm c}$ = 0.3, see Col.\,4. Accounting for the maximal frequency dependence of $f_{\rm c}$, 
$f_{\rm c}$ = 0.2 $\times$ ($\nu$/26\,GHz)$^{0.5}$, the corresponding reduced $\chi^2$ values are given in the last column. 
A comparison of the results summarized in the last two columns provides a good measure of the uncertainties involved. 
\end{tablenotes}
\end{flushleft}
\end{threeparttable}
\end{table}

Instead of applying a constant $f_{\rm c}$ = 0.3, we can also use the opposite extreme, an $f_{\rm c}$ value with the strongest 
frequency dependence permitted by the boundary conditions discussed in Sect.\,4.1. For $f_{\rm c}$ = 0.2 $\times$ 
($\nu$/26\,GHz)$^{0.5}$ the resulting optical depths are given in the right column of Table~3. With $f_{\rm c}$ decreasing 
with decreasing frequency, the lower HC$_3$N transitions become more prominent ($\tau_{\rm real}$ $\propto$ $f_{\rm c}^{-1}$) so 
that less excitation is required to simulate the calculated optical depths. The right column of Table~4 indicates this change. 
Excluding collisions, the lowest reduced $\chi^2$ is reached at $T_{\rm CMB}$ = 10\,K instead of the previously obtained 20\,K. 
With $T_{\rm CMB}$ = 5.14\,K and introducing collisions, we get $n$(H$_2$) $\sim$ 1700\,cm$^{-3}$ instead of 2600\,cm$^{-3}$. 
Again, calculated and measured line opacities differ by about $\pm$15\%. Overall, the $\chi^2_{\rm red}$ values tend to 
be smaller than those obtained with $f_{\rm c}$ = 0.3. This may indicate that the approach with variable $f_{\rm c}$ is the better one. 

To compare our results with those for the slightly less redshifted source B0218+357 (Henkel et al. 2005, 2008; Jethava et al. 
2007), we conclude that {\it in PKS1830--211 kinetic temperature and gas density are much higher, while the source covering
factor may be less frequency dependent}. However, the situation may not be quite as clear as it appears at first sight. The
kinetic temperatures were measured with the same tool, namely the inversion lines of ammonia, thus providing an objective 
comparison. The gas densities, however, are determined in different ways, analyzing formaldehyde (H$_2$CO) in the case of 
B0218+357 and cyanoacetylene (HC$_3$N) in the case of PKS\,1830--211. This may lead to a systematic bias. If we were able to 
detect HC$_3$N absorption in its ground rotational transition toward PKS\,1830--211, for example, the density of the gas traced 
by this line should only be a few 100\,cm$^{-3}$ and not 2000\,cm$^{-3}$. Otherwise the level populations would be inverted. 
While HC$_3$N appears to be very weak in B0218+357, an H$_2$CO study like that of Jethava et al. (2007) for B0218+357 is also 
feasible for PKS\,1830-211 to provide a comparison of densities excluding any potential bias caused by the use of different 
molecular tracers.

In view of our goal to constrain $T_{\rm CMB}$ (see Sects.\,4.1.2 and 4.1.3), it is important that the density derived is not
sensitively dependent on the choice of the temperature of the cosmic microwave background. To test this, we also calculated 
models with $T_{\rm CMB}$ = 4\,K and 7\,K instead of 5.14\,K. Resulting densities are $n$(H$_2$) $\sim$ 2000\,cm$^{-3}$ and 
$\sim$1000\,cm$^{-3}$ instead of $\sim$1700\,cm$^{-3}$, when using frequency dependent source covering factors.

\begin{table*}
\label{tab5}
\begin{threeparttable}
\caption[]{Integrated optical depths of optically thin lines (see Sects.\,4.1.2 and 4.1.3) and resulting
excitation temperatures.}
\begin{flushleft}
\begin{tabular}{ccccccc}
\hline
Species   & \multicolumn{3}{c}{$\int$$\tau$\,d$V$} & \multicolumn{3}{c}{Excitation temperature} \\
          & $J$=1$\leftarrow$0 & $J$=2$\leftarrow$1$^a$ & $J$=2$\leftarrow$1/$J$=1$\leftarrow$0$^a$ &
  $T_{\rm ex, 1-0}^b$ & $T_{\rm ex, 2-1}^{a,b}$ & $T_{\rm ex,LTE}^c$ \\
          &          \multicolumn{2}{c}{(km\,s$^{-1}$)}   &                                             &
  \multicolumn{3}{c}{(K)}                                            \\
\hline
                                &                 &                  &      &      &             \\
   SiO        & 0.412$\pm$0.021 & 1.440$\pm$0.270 &  3.495$\pm$0.679 & 5.51 & 5.21 & 23          \\
C$^{34}$S$^a$ & 0.292$\pm$0.020 & 0.440$\pm$0.040 &  1.507$\pm$0.172 & 5.83 & 5.27 & 7.2$\pm$0.4 \\
H$^{13}$CO$^+$& 2.804$\pm$0.259 & 2.500$\pm$0.050 &  0.892$\pm$0.084 & 5.20 & 5.17 & 3.8$\pm$0.3 \\
H$^{13}$CN    & 1.791$\pm$0.243 & 2.100$\pm$0.080 &  1.173$\pm$0.165 & 5.37 & 5.24 & 4.8$\pm$0.5 \\
HN$^{13}$C    & 0.307$\pm$0.181 & 0.900$\pm$0.070 &  2.932$\pm$1.743 & 5.20 & 5.17 &     --      \\
                                &                 &                  &      &      &             \\
CS-NE$^d$     & 0.080$\pm$0.010 & 0.280$\pm$0.080 &  3.500$\pm$1.092 &  --  &  --  & 11$\pm$3    \\
\hline
\end{tabular}
\begin{tablenotes}
\item[a)] In the cases of C$^{34}$S and CS, $J$=2$\leftarrow$1 should be replaced by $J$=4$\leftarrow$3. With the exception 
of the SiO $J$=2$\leftarrow$1 line (Fig.\,\ref{fig3}), the $J$=2$\leftarrow$1 and 4$\leftarrow$3 data were taken from Muller 
et al. (2006). 
\item[b)] LVG excitation temperatures of optically thin lines with $T_{\rm kin}$ = 80\,K, $T_{\rm CMB}$ = 5.14\,K,
and $n$(H$_2$) = 2000\,cm$^{-3}$.
\item[c)] Excitation temperatures under conditions of Local Thermodynamical Equilibrium (LTE). For the uncertainty 
in the estimate from SiO (this includes so far unpublished SiO $J$=4--3 data), see Sect.\,4.1.2.
\item[d)] CS absorption toward the north-eastern continuum component, displaced from the main absorption line by 
--147\,km\,s$^{-1}$. Since no spatial density has been derived for this feature, Cols.\,5 and 6 are left empty.
\end{tablenotes}
\end{flushleft}
\end{threeparttable}
\end{table*}

\subsubsection{SiO}

HC$_3$N was particularly interesting, because it is a heavy molecule with a small rotational contant. In PKS\,1830--211, 
HC$_3$N rotational lines can be observed at multiples of $\nu$ $\sim$ 4.8\,GHz, allowing us to detect several transitions. 
Small frequencies lead to slow rates of spontaneous decay ($A_{\rm ij}$ $\propto$ $\nu^3$) and level populations are already 
altered by collisional processes in a rather tenuous interstellar medium with $n$(H$_2$) $\sim$ 1000\,cm$^{-3}$. 

There exist a large number of lighter interstellar molecules with higher rotational constants and low apparent optical
depths, indicating as in the case of HC$_3$N optically thin absorption. SiO is one of these species, which exemplifies some 
effects we have to account for in a thorough analysis of PKS\,1830--211. SiO has a five times larger rotational constant 
than HC$_3$N and a similar electric dipole moment. With $T_{\rm CMB}$ = 5.14\,K and a density of 2000\,cm$^{-3}$ (Sect.\,4.1
and Table~4), LVG modeling (collision rates from Turner et al. 1992) yield in the optically thin limit excitation 
temperatures of 5.51\,K and 5.21\,K for the SiO $J$=1$\leftarrow$0 and 2$\leftarrow$1 lines, respectively. This implies that 
the densities derived in Sect.\,4.1.1 are not drastically modifying line intensities and that, if SiO arises from the same gas 
as HC$_3$N, the bulk of the excitation is provided by the CMB. 

Line-to-continuum ratios of 0.13 or 0.38 are observed in the CS 1$\leftarrow$0 (Fig.\,\ref{fig1}) or HCN and HCO$^+$ 
$J$=2$\leftarrow$1 lines (Muller \& Gu{\'e}lin 2008), providing firm boundary conditions for the frequency range from the 
SiO 1$\leftarrow$0 to the 2$\leftarrow$1 transition. Since the SiO lines are much weaker, they must be optically thin. 
With integrated apparent optical depths of (0.412$\pm$0.021)\,km\,s$^{-1}$ and (1.44$\pm$0.27)\,km\,s$^{-1}$ for the 1$\leftarrow$0 
and 2$\leftarrow$1 lines, respectively, we obtain $\tau$(2$\leftarrow$1)/$\tau$(1$\leftarrow$0) = 3.50$\pm$0.68 (Table~5). 
Assuming Local Thermodynamical Equilibrium (LTE), this results in an excitation temperature of 23\,K, with the expected 5\,K 
(requiring a ratio of 2.2) at a 2$\sigma$ level. We conclude that the noise in the SiO $J$=2$\leftarrow$1 line is still too high 
for a definite result. More sensitive $J$=2$\leftarrow$1 data are therefore required. 

Using SiO as an example, we should also note that there exists the possibility that the absorption exclusively arises from 
a much denser gas component than HC$_3$N. Unlike emission lines, the strength of absorption lines does not depend on 
``critical densities'', where timescales of radiative de-excitation match those for collisional excitation. Therefore, a low 
excitation level caused by large frequencies and dipole moments does not necessarily lead to non-detections and a high 
ratio of the optical depths between the 2$\leftarrow$1 and 1$\leftarrow$0 lines of SiO at $T_{\rm CMB}$ $\sim$ 5.14\,K could 
be the consequence of chemical differentiation. A test with a more recently taken SiO $J$=4--3 line ($\int{\tau}$\,d$v$ = 
0.70$\pm$0.07; S. Muller, priv. comm.) yields $T_{\rm rot}$ = (6.8$\pm$0.3)\,K. While this may indicate that there is no dense
SiO component, we emphasize that half a decade lies between this recent measurement from August/September 2006 and the 
other data presented here. Variability of the source might thus lead to a larger uncertainty than the given formal error. 

Tests with our LVG model show another effect. The excitation of the $J$=1$\leftarrow$0 line rises with increasing density, 
becoming inverted at $n$(H$_2$) $\sim$ 6$\times$10$^4$\,cm$^{-3}$. While rising, the $J$=1$\leftarrow$0 optical depth 
decreases, keeping the product $\tau$\,$T_{\rm ex}$ approximately constant. Thus $\tau$(2$\leftarrow$1)/$\tau$(1$\leftarrow$0) 
ratios in excess of four (the maximum for LTE conditions) are possible and may become a sensitive probe of densities 
related to molecular constituents, unaffected by critical densities. For $T_{\rm kin}$ = 80\,K and $T_{\rm CMB}$ = 5.14\,K, 
a density of 1.3$\times$10$^{4}$\,cm$^{-3}$ would reproduce the measured but uncertain ratio of $\sim$3.5. This density is 
almost an order of magnitude higher than that derived from HC$_3$N.

\subsubsection{Other molecular species} 

Here we will analyze opacity ratios between optically thin $J$=1$\leftarrow$0 lines measured by us (Fig.\,\ref{fig3}) and 
the corresponding $J$=2$\leftarrow$1 or $J$=4$\leftarrow$3 lines previously reported by Muller et al. (2006; their Table~4). 
CS has a rotational constant similar to that of SiO. HCN, HCO$^+$, and HNC have rotational constants about twice as large 
and an order of magnitude larger than that of HC$_3$N. This drastically reduces effects caused by collisional excitation 
as long as these species arise from the same volume as HC$_3$N. Columns 5 and 6 of Table~5 show this effect. With the 
boundary conditions of $T_{\rm kin}$ = 80\,K, $T_{\rm CMB}$ = 5.14\,K, and $n$(H$_2$) = 2000\,cm$^{-3}$ (Sect.\,4.1.1), excitation 
temperatures for HCN, HCO$^+$, and HNC are only slightly larger than $T_{\rm CMB}$, while for CS the difference is a little
larger. The values were obtained from LVG calculations with collision rates of Turner et al. (1992), Flower (1999), and 
Sch{\"o}ier et al. (2005).

While the signal-to-noise ratio of our HN$^{13}$C $J$=1$\leftarrow$0 line is not high enough to derive a meaningful value, 
the average of the LTE excitation temperatures of C$^{34}$S, H$^{13}$CO$^+$, and H$^{13}$CN is close to the expected value of 
5.14\,K.  However, deviations of individual values (see Col.\,7 of Table~5) are significant in most cases and are well 
beyond the estimated uncertainties. Combes \& Wiklind (1999) showed a diagram of previously determined LTE temperatures 
(their Fig.\,4), obtained with the SEST 15-m and IRAM 30-m telescopes. These results also show significant scatter. While LTE 
temperatures higher then the expected $T_{\rm CMB}$ value could be explained by additional collisional or radiative excitation 
(in the case of our C$^{34}$S results, a density of $n$(H$_2$) = 10$^4$\,cm$^{-3}$ would be needed), those ratios below the 
expected $T_{\rm CMB}$ value, obtained by Combes \& Wiklind (1999) and by us from linear molecules, are more difficult to 
interpret. Combes \& Wiklind (1999) suggested that their low values, not including $J$=1$\leftarrow$0 transitions, are caused 
by the heating of the 1$\leftarrow$0 and an associated cooling of the 2$\leftarrow$1 transition. Our data do not confirm 
this hypothesis, indicating instead that the scatter in the LTE excitation temperatures is much higher than the errors in 
Col.\,7 of Table~5 suggest. A dedicated effort including many more species, among them also asymmetric tops, may solve this 
problem. So far, the molecular absorption lines toward PKS\,1830--211 do not provide the kind of constraints on $T_{\rm CMB}$, 
which one is also attempting to obtain in analyzing atomic fine-structure levels in redshifted sources (e.g., Ge et al. 1997; 
Srianand et al. 2000; Molaro et al. 2002).

\subsubsection{The northwestern source}

As already pointed out in Sect.\,3, absorption toward the south-western compact continuum component of PKS\,1830--211 is 
relatively strong. However, a few prominent molecular lines are also detected at a velocity offset of --147\,km\,s$^{-1}$ 
toward the north-eastern component, which is located behind another spiral arm of the lensing galaxy (e.g., Wiklind \& Combes 1998; 
Muller et al. 2006). While our HCO$^+$ profile (Fig.\,\ref{fig1}) is not calibrated well enough for an analysis, the CS 
$J$=1$\leftarrow$0 feature (also Fig.\,\ref{fig1}) can be compared with the corresponding $J$=4$\leftarrow$3 profile of Muller 
et al. (2006). The resulting LTE excitation temperature is given in the last line of Table~5. As a caveat, however, we note that 
the CS spectrum from Muller et al. (2006) was taken in July 1999, more then two years before our measurements (see Muller \& 
Gu{\'e}lin 2008 for source variability with time). This time interval is longer than for the spectra from the south-western 
source.

\begin{table}
\label{tab6}
\begin{threeparttable}
\caption[]{Radial velocities (see Sect.\,4.2). Uncertainties in the frequencies are small relative to those of the velocities, 
which were obtained by Gaussian fits$^{a}$.}
\begin{flushleft}
\begin{tabular}{crrr}
\hline 
\multicolumn{1}{c}{Species} & \multicolumn{1}{c}{Transition} & \multicolumn{1}{c}{Rest}       & \multicolumn{1}{c}{$V_{\rm LSR}$}  \\
                            &                                & \multicolumn{1}{c}{Frequency}  & \multicolumn{1}{c}{(km\,s$^{-1}$)} \\
                            &                                & \multicolumn{1}{c}{(GHz)}      &                                    \\
\hline
              &                        &            &                \\
   NH$_3$     &  ($J$,$K$) = (1,1)     & 23.694496  &    9.0\,(0.1)  \\
              &                        &            &    8.6\,(0.5)  \\
              &              (2,2)     & 23.722633  &    7.4\,(0.2)  \\
              &                        &            &   10.5\,(0.4)  \\
              &              (3,3)     & 23.870130  &    8.6\,(0.1)  \\
              &              (4,4)     & 24.139416  &    8.4\,(0.2)  \\
              &                        &            &    9.1\,(0.7)  \\
              &              (5,5)     & 24.532989  &    8.2\,(0.3)  \\
              &                        &            &    8.7\,(0.5)  \\
              &              (6,6)     & 25.056025  &    8.1\,(0.2)  \\
              &                        &            &    8.5\,(0.6)  \\
              &              (7,7)     & 25.715182  &   10.9\,(0.4)  \\
              &              (8,8)     & 26.518981  &    8.5\,(0.6)  \\
              &              (9,9)     & 27.477943  &    8.8\,(0.7)  \\
              &            (10,10)     & 28.604737  &   10.2\,(2.0)  \\
              &                        &            &                \\
HC$_3$N       &  $J$ = 3$\leftarrow$2  & 27.294289  &    9.0\,(0.4)  \\
              &        4$\leftarrow$3  & 36.392324  &    8.2\,(0.3)  \\
              &        5$\leftarrow$4  & 45.490314  &    8.0\,(1.0)  \\
              &        6$\leftarrow$5  & 54.588247  &    7.7\,(0.3)  \\
              &        7$\leftarrow$6  & 63.686052  &   10.0\,(0.4)  \\
              &                        &            &                \\
SiO           &        1$\leftarrow$0  & 43.423846  &    9.8\,(0.7)  \\
SiO           &        2$\leftarrow$1  & 86.847010  &    8.9\,(1.8)  \\
C$^{34}$S     &        1$\leftarrow$0  & 48.206946  &    9.8\,(0.6)  \\
H$^{13}$CO$^+$&        1$\leftarrow$0  & 86.754288  &    9.7\,(0.7)  \\
H$^{13}$CN    &        1$\leftarrow$0  & 86.339922  &    8.8\,(1.5)  \\ 
HN$^{13}$C    &        1$\leftarrow$0  & 87.090873  &    7.4\,(1.3)  \\ 
\hline
\end{tabular}
\begin{tablenotes}
\item[a)] Frequencies were taken from the JPL catalog (NH$_3$), the Cologne
Database for Molecular Spectroscopy (HC$_3$N), and the line list of Lovas
(other species). NH$_3$ transitions with two $V_{\rm LSR}$ values were 
measured twice (see Henkel et al. 2008).
\end{tablenotes}
\end{flushleft}
\end{threeparttable}
\end{table}

\subsection{The proton-to-electron mass ratio}

\subsubsection{Some general considerations}

As already mentioned in Sect.\,1, a comparison of radial velocities from inversion lines of ammonia (NH$_3$) with rotational 
lines of either NH$_3$ or other species can provide sensitive constraints on variations in the proton-to-electron mass ratio 
$\mu$ = $m_{\rm p}$/$m_{\rm e}$. Following Flambaum \& Kozlov (2007), 
\begin{equation}\label{eq:mu}
  \frac{\Delta V}{\rm c} = \frac{z_{\rm inv} - z_{\rm rot}}{1 + z} = 3.46 \times \ \frac{\Delta \mu}{\mu},
\end{equation}
with $z_{\rm inv}$ and $z_{\rm rot}$ denoting the apparent redshifts of the inversion and rotational lines, $z$ representing the 
real redshift of the source, and $\Delta$$\mu$ providing the deviation relative to the current laboratory value of $\mu$, defined 
such that a negative value indicates a smaller $\mu$ in the absorbing cloud. In the case where both the NH$_3$ and rotational 
transitions show absorption in just a single velocity component, defining and measuring the velocity difference $\Delta V$ is 
straightforward. The more general case where many velocity components exist must be treated differently, by varying all the free 
parameters, including $\Delta V$, to minimize $\chi^2$ in a simultaneous fit to all NH$_3$ and rotational profiles (e.g., Murphy 
et al. 2008a).

Ammonia and cyanoacetylene (HC$_3$N) are both molecular high density tracers. Toward PKS\,1830--211, the NH$_3$ inversion lines 
and the rotational HC$_3$N lines are optically thin and observed at similar frequencies. Therefore, in this paper we derive 
$\Delta\mu/\mu$ by comparing the NH$_3$ and HC$_3$N spectra. Given the fairly low signal-to-noise ratio of these spectra, there is 
little obvious evidence for more than a single absorbing velocity component across the NH$_3$ and HC$_3$N profiles. We therefore 
proceed with a simple analysis assuming that only one absorbing velocity component exists, deferring a simultaneous $\chi^2$ minimization 
analysis which incorporates other weak components and other rotational transitions to a later paper. Despite this assumption of a 
single velocity component, judicious treatment of the uncertainties on its velocity in the NH$_3$ and rotational transitions will 
ensure our constraint on $\Delta\mu/\mu$ is robust, if not optimally precise. 

The NH$_3$ inversion lines were measured with the Green Bank Telescope (GBT), while the HC$_3$N data were obtained
at Effelsberg. The Local Standard of Rest velocity scales at both sites are, however, well adjusted (Tifft \& Huchtmeier 
1990) and have not been significantly modified in recent years. They refer to a peculiar solar velocity of 20\,km\,s$^{-1}$ 
toward $\alpha_{1900}$ = 18$^{\rm h}$ and $\delta_{1900}$ = +30$^{\circ}$, corresponding to $V_{\rm LSR}-V_{\rm HEL}$ = 12.432\,km\,s$^{-1}$ 
in the case of PKS\,1830--211. Therefore, the velocity measurements listed in Table~6 can be directly compared to estimate $\Delta\mu/\mu$ 
and its uncertainty, without further velocity corrections.

\subsubsection{Average velocities of NH$_3$ and HC$_3$N}

For the single-Gaussian fits to the NH$_3$ transitions, Table~6 reveals somewhat inconsistent centroid velocities. They scatter 
around the weighted mean velocity over all transitions of 8.65\,km\,s$^{-1}$ with a $\chi^2$ value of 117 which, for just 14 
degrees of freedom, is far more than expected based on the individual velocity uncertainties. This may indicate the presence of 
more than one velocity component. Despite the large range of NH$_3$ excitation states observed by Henkel et al. (2008) -- some 
1030\,K -- we do not observe a trend in velocity with excitation state: For the three lowest NH$_3$ inversion lines, the unweighted 
mean velocity is 8.82$\pm$0.45\,km\,s$^{-1}$ and for the higher excited ones we find a mean of 8.94$\pm$0.27\,km\,s$^{-1}$ 
(the errors are standard deviations on the means). Indeed, the larger than expected scatter in fitted velocities for the different 
NH$_3$ transitions seems to be random in sign and magnitude. A similar additional scatter is also observed in the rotational HC$_3$N 
transitions (Fig.\,\ref{fig2}). These cover only a small amount of excitation, with the highest level, $J$=7, being just $\sim$12\,K 
above the ground state.

Given the velocity scatter in the transitions of NH$_3$ and HC$_3$N, constraining the variation of $\mu$ is best done with conservative
velocity and uncertainty estimates which naturally incorporate the observed scatter. The simplest such velocity estimator is obviously
the unweighted mean velocity and its standard deviation which, for the NH$_3$ transitions in Table~6, are 8.90$\pm$0.24\,km\,s$^{-1}$, 
and for the HC$_3$N transitions, are 8.58$\pm$0.37\,km\,s$^{-1}$. Of course, the statistical velocity uncertainties quoted in Table~6 
for some transitions are so high that including them in the mean velocity calculation is likely to decrease the reliability of the 
mean. We therefore reject transitions with velocity uncertainties larger than the root-mean-square (RMS) velocity variation for each 
species, 0.91 and 0.83\,km\,s$^{-1}$ for NH$_3$ and HC$_3$N respectively. With this criterion, the NH$_3$ (10,10) and HC$_3$N 
5$\leftarrow$4 transitions are rejected. {\it The final mean velocity for the remaining NH$_3$ transitions is 8.81$\pm$0.23\,km\,s$^{-1}$ 
and, for the remaining HC$_3$N transitions, is 8.73$\pm$0.43\,km\,s$^{-1}$}.

\subsubsection{Resulting limits for a variation of $\mu$}

Using these final clipped mean velocities, $\Delta V$ = 0.08$\pm$0.49\,km\,s$^{-1}$. Equation \ref{eq:mu} then provides our 
1-$\sigma$ constraint on the variation in $\mu$, $\Delta\mu/\mu$ = (+0.08$\pm$0.47) $\times$ 10$^{-6}$. Since the quoted uncertainty 
derives entirely from the scatter in the individual transition velocities, this should be a reasonably robust error estimate. 
Nevertheless, given that we have used only single Gaussian fits to the absorption profiles and that there is some scatter in the 
individual transition velocities, we quote our final result as a 3-$\sigma$ upper limit on variation in $\mu$ at the absorption 
redshift of $z=0.89$,
\begin{equation}\label{eq:dmu_z} 
\left|\Delta\mu/\mu\right| < 1.4\times10^{-6}\,.
\end{equation}
For comparison with laboratory constraints on variations in $\mu$, and in the absence of a reliable model for how $\mu$ might be 
expected to vary with cosmological time, it is common, if not well motivated, to assume that any variation is linear in time. Hence, 
our upper limit on variation in $\mu$ translates to a 3-$\sigma$ upper limit on its time variation 
\begin{equation}\label{eq:dmu_t}
\left|\dot{\mu}/\mu\right| < 2.0 \times\ 10^{-16}\,{\rm yr}^{-1}
\end{equation}
over the past 7.0 Gyr.

There is yet another study on the proton-to-electron mass ratio in the main lens of PKS\,1830--211. From a comparison of the 
ammonia inversion lines (Henkel et al. 2008) with its ($J$,$K$) = (1,0)$\leftarrow$(0,0) rotational transition, Menten et al. 
(2008) find consistency, at a 1$\sigma$ level, of $\Delta\mu$/$\mu$ = 1.9$\times$10$^{-6}$. The strength of this study is its 
focus on lines with different dependencies on $\mu$ arising from the same molecular species. On the other hand, the ratio between 
the frequencies of the rotational and inversion lines is $\sim$25, thus leading to potentially significant differences in the 
morphology and the covering factor of the background radio and submillimeter continuum.  While such systematic differences 
cannot be quantified on the basis of a single rotational line, the given uncertainty of the resulting $\Delta\mu$/$\mu$ value 
is dominated by the limited signal-to-noise ratio of the rotational line.

\subsubsection{Potential caveats}

The consistency of NH$_3$ and HC$_3$N radial velocities is remarkable in view of a number of effects, which might exert a 
significant influence on our results. Having carefully avoided the use of optically thick transitions, these are (1) time 
variability of the continuum source, (2) a frequency dependent continuum morphology, (3) hyperfine structure, (4) chemistry,
and (5) inhomogeneities in temperature and density inside a cloud of size $\ga$10\,pc (Carilli et al. 1998). 

(1) A time variable continuum source may lead to different lines-of-sight and thus to different radial velocities at different 
epochs. Not much of this was seen in spite of the existence of monitoring programs (Wiklind \& Combes 1998; Muller et al. 2006;
Muller \& Gu{\'e}lin 2008) or repeated measurements of ammonia lines (Table~6 and Henkel et al. 2008). 

(2) The NH$_3$ and HC$_3$N transitions used in this study are much closer in frequency than those chosen by Murphy et al. 
(2008a) and Menten et al. (2008). Also, in PKS\,1830--211, the source covering factor may be less frequency dependent than in 
B0218+357 (Sect.\,4.1.1). Nevertheless, NH$_3$ and HC$_3$N frequencies differ by factors of 1.0--2.5. While no direct effect 
is apparent, it remains a source of uncertainty which cannot be quantified. 

(3) In particular NH$_3$ velocities would be greatly affected by hyperfine (hf) splitting, if strong non-LTE effects would 
occur. A main group of hf-components is surrounded by four satellite groups, displaced by about $\pm$10 and $\pm$20\,km\,s$^{-1}$. 
The higher the energy above the ground state of an inversion line, the weaker the satellite features relative to the main
group (e.g., the satellites account for $\sim$2.5\% of the total absorption in the ($J$,$K$) = (7,7) line). The above made 
comparison of average velocities of the three lowest with the seven higher excited NH$_3$ inversion lines does not show a significant 
shift in velocity. Deviations from LTE intensity ratios of the various components are only expected in the case of a significant 
population of the non-metastable inversion states ($J$$>$$K$), which requires extremely high densities or intense radiation fields 
(Stutzki et al. 1984; Stutzki \& Winnewisser 1985a,b). Non-LTE effects leading to HC$_3$N velocity shifts as a function of rotational 
quantum number $J$ are also not obvious. In this case, the $J$=3$\leftarrow$2 line would be the most critical, while the hf-structure 
of the 7$\leftarrow$6 line is far too compact to yield any significant shifts.

(4) Chemistry is a more difficult issue, because there is no way to estimate its effect on our data. NH$_3$ and HC$_3$N may 
have different spatial distributions, which is most impressively exemplified by the nearby prototypical dark cloud TMC-1, where 
peaks of line emission are offset by 7\,arcmin (e.g., Olano et al. 1988). HC$_3$N is a molecule representing a young chemical 
age, while NH$_3$ refers to a chemically more evolved stage of cloud chemistry (e.g., Dickens et al. 2001; Hirota et al. 2002). 

(5) Since NH$_3$ is covering an enormous range of excitation without showing any significant change in radial velocity, changes 
in temperature and density may also not play an important role.

\subsubsection{Summary and outlook}

To summarize, we either have the unlikely scenario in which different effects cancel each other out, or all of these effects are 
minor. In order to broaden the scope of our study and to avoid relying entirely on the rotational transitions of HC$_3$N, we may 
also use the $J$=1$\leftarrow$0 and 2$\leftarrow$1 lines presented in Fig.\,\ref{fig3}. We should note, however, that for these 
lines, frequencies are significantly higher than for NH$_3$ and HC$_3$N. {\it The unweighted mean rotational velocity then becomes 
9.07$\pm$0.38\,km\,s$^{-1}$}. The {\it velocity difference} amounts to {\it $\Delta V$ = +0.26$\pm$0.44\,km\,s$^{-1}$ relative 
to NH$_3$, while for HC$_3$N we have found --0.08$\pm$0.49\,km\,s$^{-1}$}. These differences highlight potential systematic effects, 
caused by a time variable continuum background, by a frequency dependent continuum morphology, by non-LTE effects in lines with 
hyperfine structure, by astrochemistry, and by the specific excitation requirements of individual lines. The large number of 
analyzed molecular transitions, covering a wide parameter space, ensures that systematic discrepancies in velocity are within 
1\,km\,s$^{-1}$, corresponding to $\Delta \mu$/$\mu$ $<$10$^{-6}$.

Our constraint on variations of the proton-to-electron mass ratio $\mu$ differs in several ways from that given by Murphy et al. (2008a) 
for B0218+357 at $z$$\sim$0.68 (this study was also based on a comparison of ammonia (NH$_3$) inversion lines with rotational transitions 
of other molecular species). (1) The frequencies of the lines measured by us are much more similar than in the previous study, minimizing 
effects caused by a frequency dependent continuum background morphology on the observed column. (2) All our analyzed lines are optically 
thin, avoiding modified lineshapes due to saturation and (3) our number of observed transitions is much larger. This may result 
in an evaluation of hidden systematic errors, which is more realistic than any estimate from multi-velocity component fits of a few 
spectral lines only. (4) The time interval between our measurements of inversion and rotational lines is smaller, possibly reducing 
systematic effects caused by a time variable radio continuum background. Note, however, that this may be compensated by a higher 
variability of our source. 

There are also two weaknesses with respect to the study of Murphy et al. (2008a). (1) The excitation of most inversion lines analyzed by 
us is much higher than that of the rotational lines. While no trend of radial velocities as a function of excitation is seen by us, the 
study of Murphy et al.  (2008a) is confined to inversion lines of modest excitation and is definitely less sensitive to potential 
discrepancies with respect to excitation. (2) We did not yet provide a multi-component spectral analysis of the entire dataset, 
accounting for velocity structure and hyperfine splitting. This is {\it the} missing step toward a complete analysis of the spectra 
and will be discussed in a forthcoming paper, focusing entirely on potential variations of $\mu$.

\section{Conclusions}

Our observations of molecular lines at redshift $z$$\sim$0.89 toward the gravitational lens system PKS\,1830--211 reveal the 
following main results:

\begin{itemize}

\item Cyanoacetylene (HC$_3$N) is detected in absorption in five rotational lines. The $J$=7$\leftarrow$6 line may have been 
detected for the first time in the interstellar medium. CS, C$^{34}$S, HCO$^+$, HC$^{13}$O$^+$, H$^{13}$CN, and SiO absorption 
are also detected, in their ground rotational transitions. 

\item A Large Velocity Gradient analysis of the HC$_3$N spectra yields, adopting $T_{\rm kin}$ = 80\,K from ammonia (NH$_3$),
a density of $n$(H$_2$) $\sim$ 2600\,cm$^{-3}$ for a background source covering factor $f_{\rm c}$, which is independent of 
frequency. CS, HCO$^+$, and OH data constrain its frequency dependence to $f_{\rm c}$ $\propto$ $\nu^{0.0...0.5}$. This holds 
for the frequency range between 26\,GHz and 94\,GHz. For a maximal $f_{\rm c}$ exponent of 0.5, $n$(H$_2$) is reduced to 
$\sim$1700\,cm$^{-3}$.

\item The gas density determined using HC$_3$N is low enough that molecules with higher rotational constants should be mainly 
excited by the cosmic microwave background. Therefore, we compare our C$^{34}$S, H$^{13}$CO$^+$, and H$^{13}$CN $J$=1$\leftarrow$0 
data with the corresponding $J$=2$\leftarrow$1 lines ($J$=4$\leftarrow$3 in the case of C$^{34}$S) presented by Muller et al. 
(2006). While the average excitation temperature is consistent with the expected temperature of the microwave backgrund, 
$T_{\rm CMB}$ = 5.14\,K, deviations of individual values (3--7\,K) far exceed expected uncertainties. 

\item For the weak north-eastern absorption component, a comparison of the CS $J$=1$\leftarrow$0 and 4$\leftarrow$3 transitions
yields an excitation temperature of (11$\pm$3)\,K. The comparison is, however, based on data, which were taken at epochs more 
than two years apart. 

\item From a comparison of the radial velocities of five rotational transitions of HC$_3$N with ten inversion transitions of ammonia,
the proton-to-electron mass ratio $\mu$ of the lens at $z$=0.89 deviates by $\left|\Delta\mu/\mu\right| < 1.4\times10^{-6}$ from
the local value with 3-$\sigma$ (99.7\%) confidence. Assuming that any variation in $\mu$ evolves linearly with cosmological time, this
corresponds to $\left|\dot{\mu}/\mu\right| < 2.0\times10^{-16}$\,yr$^{-1}$. The large number of analyzed optically thin molecular 
transitions, covering a wide range of physical and chemical parameters, indicates that systematic effects should be $<$1\,km\,s$^{-1}$, 
corresponding to an uncertainty of $<10^{-6}$ in $\Delta\mu/\mu$.

\item Comparing the main lens of PKS\,1830-211 with that of B0218+357, the molecular gas toward PKS\,1830-211 is warmer and appears to 
be denser, while its molecular source covering factor may be slightly less frequency dependent at cm- and mm-wavelengths.

\end{itemize}

\begin{acknowledgements}
It is a pleasure to thank J.N. Chengalur, S. Thorwirth, and C.M. Walmsley for useful discussions and critical reading of the 
manuscript. MTM thanks the Australian Research Council for a QEII Fellowship (DP0877998). We used NASA's Astrophysical Data System (ADS), 
the Cologne Database for Molecular Spectroscopy (CDMS; see M{\"u}ller et al. 2001, 2005), the JPL Catalog 
(http://spec.jpl.nasa.gov/ftp/pub/catalog/catform.html), and the line lists of Lovas (1992) and Coudert \& Roueff (2006). 
\end{acknowledgements}

\end{document}